 \documentclass[10pt,twocolumn,twoside]{IEEEtran}  

\IEEEoverridecommandlockouts                              
\usepackage{fancyhdr,amssymb,amsmath, graphicx, listings,float,subfig,enumerate,epstopdf,color,setspace,bm,cite,setspace,cite}

\def\Var{\operatorname{ Var }}

\def\Pr{\operatorname{ Pr }}
\newcommand{\aee}{{\cal A}}

\newcommand{\E}{\mathbb{E}}
\newcommand{\p}{\bar{p}}
\newcommand{\s}{s}
\newcommand{\m}[1]{\boldsymbol{ #1 }}
\newcommand{\tb}[1]{\textbf{#1}}
\newcommand{\eps}{\varepsilon}
\newcommand{\given}{\,|\,}
\newcommand{\st}{\,:\,}
\DeclareMathOperator*{\argmax}{arg\,max}

\renewcommand{\P}\phi

\newtheorem{theorem}{Theorem}

\newtheorem{claim}{Claim}
\newtheorem{lemma}{Lemma}

\newtheorem{defn}{Definition}
\newtheorem{example}{Example}

\normalsize\title{\LARGE \bf
Learning Efficient Correlated Equilibria\thanks{This research was supported by AFOSR grants \#FA9550-12-1-0359, ONR grant \#N00014-12-1-0643, NSF grant \#ECCS-1351866, the NASA Aeronautics scholarship program, the Philanthropic Educational Organization, and the Zonta International Amelia Earhart fellowship program.}}

\author{
Holly P. Borowski\thanks{H. Borowski is with the Department of Aerospace Engineering Sciences,  University of Colorado, 429 UCB, Boulder, CO 80309, 719-213-3254. {\texttt{holly.borowski@colorado.edu}.}}, 
Jason R. Marden\thanks{J. R. Marden is with the Department of Electrical and Computer Engineering, Harold Frank Hall, Rm 5161, University of California, Santa Barbara, 93106, 805-893-2299, {\texttt{jrmarden@ece.ucsb.edu}.  Corresponding author.}}, 
and Jeff S. Shamma\thanks{J. S. Shamma is with the Department of Electrical and Computer Engineering, Georgia Institute of Technology and with the King Abdullah University of Science and Technology (KAUST), \texttt{jeff.shamma@kaust.edu.sa}.}}

\begin{document}

\graphicspath{{figures/}}
\maketitle

\thispagestyle{plain}
\pagestyle{plain}

\begin{abstract}
The majority of distributed learning literature focuses on convergence to Nash equilibria.
Correlated equilibria, on the other hand, can often characterize more efficient collective behavior than even the best Nash equilibrium.  However, there are  no existing distributed learning algorithms that converge to specific correlated equilibria.  In this paper, we provide one such algorithm which guarantees that the agents' collective joint strategy will constitute an efficient correlated equilibrium with high probability.  The key to attaining efficient correlated behavior through distributed learning involves incorporating a common random signal into the learning environment.
\end{abstract}

\section{Introduction} 
Agents' control laws are a crucial component of any multiagent system. They dictate how individual agents process locally available information to make decisions. Factors that determine the quality of a control law include informational dependencies, asymptotic guarantees, and convergence rates. 

Game theory has recently emerged as a framework for assigning agents' local control laws in a distributed system \cite{Lasaulce2011,Alpcan2010,Han2012,MacKenzie2006,Menache2011}. Here, a \emph{learning rule} dictates how each agent should revise its behavior, based on its individual objective and on available information about the surrounding environment. Significant research has been directed at deriving distributed learning rules that possess desirable asymptotic performance guarantees and convergence rates and enable agents to make decisions based on limited information.

The majority of this research has focused on attaining convergence to (pure) Nash equilibria under stringent information conditions \cite{Young2009, Frihauf2012, Foster2006, Boussaton2012, Poveda2013, Gharesifard2012}. Recently, the research focus has shifted to ensuring convergence to alternate types of equilibria that often yield more efficient behavior than Nash equilibria.  In particular, results have emerged that guarantee convergence to Pareto efficient Nash equilibria \cite{Marden2009,Pradelski2012}, potential function maximizers \cite{Blume1993, Marden2012}, welfare maximizing action profiles \cite{Marden2011, Arieli2012}, and the set of correlated equilibria \cite{Hart2000,Marden2013,Aumann1987,Foster1997}, among others.  

In most cases highlighted above, the derived algorithms guarantee (probabilistic) convergence to the specified equilibria.  However, the class of correlated equilibria has posed significant challenges with regards to this goal. Learning algorithms that converge to an efficient correlated equilibrium are desirable because optimal system behavior can often be characterized by a correlated equilibrium. Unfortunately, the aforementioned learning algorithms, such as regret matching \cite{Hart2000}, merely converge to the \emph{set} of correlated equilibria. This means that the long run behavior does not necessarily constitute -- or even approximate -- a specific correlated equilibrium at any instance of time.

We provide a distributed learning algorithm that converges to the most efficient, i.e., welfare maximizing, correlated equilibrium.  For concreteness, consider a mild variant of the Shapley game with the following payoff matrix
\begin{center}
\begin{tabular}{c|c|c|c|}
\multicolumn{1}{r}{}&
	\multicolumn{1}{c}{{L}}&
		\multicolumn{1}{c}{{M}}&
			\multicolumn{1}{c}{{R}}\\
\cline{2-4}T &1,-$\eps$&-$\eps$,1&0,0\\
\cline{2-4}{M}&0,0&1,-$\eps$&-$\eps$,1\\
\cline{2-4}{B}&-$\eps$,1&0,0&1,-$\eps$\\\cline{2-4}
\end{tabular}
\end{center}
where $\eps > 0$ is a small constant.  In this game, there are two players (Row, Column); the row player has three actions (T,M,B), and the column player has three actions (L,M,R). The numbers in the table above are the players' payoffs for each of the nine joint actions.  The unique Nash equilibrium for this game occurs when each player uses a probabilistic strategy that selects each of the three actions with probability $1/3$. This yields an expected payoff of approximately $1/3$ to each player.  Alternatively,  a joint distribution that places a mass of $1/6$ on each of the six joint actions that yield non-zero payoffs to the players yields an expected payoff of approximately $1/2$ to each player. Note that this distribution cannot be realized by independent strategies associated with the two players, but instead represents a specific correlated equilibrium.

As the above example demonstrates, distributed learning algorithms that converge to efficient correlated equilibria can be desirable from a system-wide perspective.  In line with this theme, results presented in \cite{Lim2014} rely on looking for cyclic behavior against a bounded memory opponent. Additionally, a recent result in \cite{Marden2013} proposed a distributed algorithm that guarantees that the empirical frequency of the agents' collective behavior will converge to an efficient correlated equilibrium; however, convergence in empirical frequencies is attained through deterministic cyclic behavior of the agents.   For example, in the above Shapley game, the algorithm posed in \cite{Marden2013} guarantees that the collective behavior of the agents will follow the cycle 
$ (T,L)  \rightarrow (T,M) \rightarrow (M,M) \rightarrow (M,R) \rightarrow (B,R) \rightarrow (B,L) \rightarrow (T,L)$
with high probability.  Following this deterministic cycle results in an empirical frequency of play that equates to the efficient correlated equilibrium highlighted above; however, at any time instance the players are not playing a joint strategy in accordance with this efficient correlated equilibrium.

Predictable, cyclic behavior may be desirable from a system-wide perspective for many applications, e.g., data ferrying \cite{Carfang2013}. However, such behavior could be exploited in many other situations, e.g., team versus team zero-sum games \cite{Ho1974, Stengel1997}.  By viewing each team as a single player, classical results for two-player zero-sum games suggest that a team's desired strategy is to play its security strategy, which can be characterized by a probability distribution over the team's joint action space. 
Distributed learning algorithms that can stabilize specific joint strategies, such as correlated equilibria, may be necessary for providing strong performance guarantees in such settings.  

In this paper we present a distributed learning algorithm that ensures the agents collectively play a joint strategy corresponding to the efficient correlated equilibrium.  With regards to the Shapley game, our algorithm guarantees that the agents collectively play the highlighted joint distribution with high probability.  Attaining such guarantees on the underlying joint strategy is non-trivial as we aim to ensure desired correlated behavior through the design of learning rules where individual agents make independent decisions in response to local information.  The key element of our algorithm that makes this correlation possible is the introduction of a common random signal to the agents, which is incorporated into their local decision-making rule.  Another important feature of our algorithm is that it is completely uncoupled \cite{Foster2006}, i.e., agents make decisions based only on their received utility and their observation of the common random signal.  In such settings, agents have no knowledge of the payoff or behavior of other agents, nor do they have any information regarding the structural form of their utility functions.   

It is important to highlight the recent results which focus on efficient centralized algorithms for computing specific correlated equilibria \cite{Jiang2011, Papadimitriou2005, Ortiz}.  Such algorithms often require  a complete characterization of the game which is unavailable in many engineering multiagent systems.  Hence, the applicability of such results to the design and control of multiagent systems may be limited.

\section{Background}

We consider the framework of finite strategic form games where there exists an agent set $N = \{1, 2, \dots, n\}$, and each agent $i \in N$ is associated with a finite action set $\aee_i$ and a utility function $U_i : \aee \rightarrow [0,1]$ where $\mathcal{A} = \mathcal{A}_1\times\mathcal{A}_2\times\cdots\times \mathcal{A}_n$ denotes the joint action space. 
%
%
We  represent such a game by the tuple $G = \left (N,\{U_i\}_{i\in N},\{\mathcal{A}_i\}_{i\in N}\right)$.

In this paper we focus on the class of coarse correlated equilibria \cite{Aumann1987}.  A coarse correlated equilibrium is characterized by a joint distribution $q = \{q^a\}_{a \in \aee} \in \Delta(\aee)$, where $\Delta(\aee)$ represents the simplex over the finite set $\aee$, such that for any agent $i \in N$ and action $a_i' \in \aee_i$,
\begin{equation} 
\sum_{a \in \aee} U_i(a_i, a_{-i}) q^a \geq \sum_{a \in \aee} U_i(a_i', a_{-i}) q^a, 
\end{equation}
where $a_{-i} = \{a_1, \dots, a_{i-1}, a_{i+1}, \dots, a_n\}$ denotes the collection of action of all players other than player $i$.\footnote{We will express an action profile $a \in \aee$ as $a=(a_i, a_{-i})$.}  Informally, a coarse correlated equilibrium represents a joint distribution where each agent's expected utility for going along with the joint distribution is at least as good as his expected utility for deviating to any fixed action.  We say a coarse correlated equilibrium $q^*$ is \emph{efficient} if it maximizes the sum of the expected payoffs of the agents, i.e., 
\begin{equation}
q^* \in \underset{q \in {\rm CCE}}{\arg \max} \sum_{i \in N} \sum_{a \in \aee} U_i(a) q^a,
\end{equation}  
where ${\rm CCE} \subset \Delta(\aee)$ denotes the set of coarse correlated equilibria.  It is well known that ${\rm CCE} \neq \emptyset$ for any game $G$.

This paper focuses on deriving a distributed learning algorithm that ensures the collective behavior of the agents converges to an efficient coarse correlated equilibrium. We adopt the framework of repeated one-shot games, where a static game $G$ is repeated over time and agents use observations from previous plays of the game to formulate a decision.  More specifically, a repeated one-shot game yields  a sequence of action profiles $a(0)$, $a(1)$, $\dots$, where at each time $t \in \{0,1,2, \dots\}$ the decision of each agent $i$ is chosen independently accordingly to the agent's strategy at time $t$, which we denote by $p_i(t) = \{p_i^{a_i}(t)\}_{a_i \in \aee_i} \in \Delta(\aee_i)$.

A learning rule dictates how each agent selects its strategy given available information from previous plays of the game. One type of learning rule, known as \emph{completely uncoupled} or \emph{payoff based} \cite{Foster2006}, takes on the form:
\begin{eqnarray}\label{eq:231}
p_i(t) = F_i\left(\left\{a_i(\tau), U_i(a(\tau))\right\}_{\tau = 0, \dots, t-1} \right)
\end{eqnarray}
%
Completely uncoupled learning rules represent one of the most informationally restrictive classes of learning rules since the only knowledge that each agent has about previous plays of the game is (i) the action the agent played and (ii) the utility the agent received. 

We gauge the performance of a learning rule $\{F_i\}_{i\in N}$ by the resulting asymptotic guarantees.  With that goal in mind, let $q(t) \in \Delta(\aee)$ represent the agents' collective strategy at time $t$, which is of the form
\begin{equation}\label{eq:123}
q^{(a_1, \dots, a_n)}(t) = p_1^{a_1}(t) \times \dots \times p_n^{a_n}(t) 
\end{equation}
where $\{p_i(t)\}_{i \in N}$ are the individual agent strategies at time $t$. The goal of this paper is to derive learning rules that guarantee the agents' collective strategy constitutes an efficient coarse correlated equilibrium the majority of the time, i.e., for all sufficiently large times $t$, 
\begin{equation}
{\rm Pr}\left[ q(t) \in \underset{q \in {\rm CCE}}{\arg \max} \sum_{i \in N} \sum_{a \in \aee} U_i(a) q^a \right] \approx 1.  
\end{equation}

Attaining this goal using learning rules of the form (\ref{eq:231}) is impossible because such rules do not allow for correlation between the players, i.e., the agents' collective strategies are restricted to being of form (\ref{eq:123}).  Accordingly, we modify the learning rules in (\ref{eq:231}) by giving each agent access to a common random signal $z(t)$ at each period $t \in \{0,1, \dots\}$ that is i.i.d. and drawn uniformly from the interval $[0,1]$.  Now,  the considered distributed learning rule takes the form
\begin{eqnarray}\label{eq:2321}
p_i(t) = F_i\left(\left\{a_i(\tau), U_i(a(\tau)), z(t))\right\}_{\tau = 0, \dots, t-1} \right).
\end{eqnarray}
As we show in the following section, this common signal can be used as a coordinating entity to reach collective strategies beyond the form given in (\ref{eq:123}).

\section{A learning algorithm for attaining efficient correlated equilibria}\label{s:learning algorithm}

In this section, we present a specific learning rule of the form (\ref{eq:2321}) that guarantees the agents' collective strategy constitutes an efficient coarse correlated equilibrium the majority of the time.  This algorithm achieves the desired convergence guarantees by exploiting the common random signal $z(t)$ through the use of \emph{signal-based strategies}.  

\subsection{Preliminaries}

Consider a situation where each agent $i \in N$ commits to a signal-based strategy of the form $s_i : [0,1] \rightarrow \aee_i$ which associates with each signal $z \in [0,1]$ an action $s_i(z) \in \aee_i$.  With an abuse of notation, we consider a finite parameterization of such signal-based strategies, which we refer to as \emph{strategies}, of the form $S_i = \cup_{\omega=1}^\Omega (\mathcal{A}_i)^\omega$ 
where $\Omega \geq 1$ is a design parameter identifying the granularization of the agent's possible strategies.  A strategy $s_i = (a_i^1,\ldots,a_i^\omega) \in S_i$, $\omega \leq \Omega$, defines a mapping of the form
\begin{equation}
s_i(z) = \left\{ \begin{array}{ccl} 
a_i^1 & \text{if} & z \in [0,1/\omega) \\ 
a_i^2 & \text{if} & z \in [1/\omega,2/\omega) \\
\vdots & & \vdots \\
a_i^\omega & \text{if} & z \in [(\omega-1)/\omega,1]. 
\end{array}
\right. 
\end{equation} 
These strategies divide the unit interval into at most $\Omega$ regions of equal length and associate each region with a specific action in the agent's action set.  If the agents commit to a strategy profile $s = (s_1, s_2, \dots, s_n) \in S = \prod_{i \in N} S_i$, the resulting joint strategy $q(s)=\{q^a(s)\}_{a \in \aee} \in \Delta(\aee)$ satisfies
$$ q^a(s) = \int_{0}^1 \prod_{i \in N} I \{s_i(z) = a_i\} dz $$
where $I\{\cdot\}$ is the indicator function.  Lastly, the set of joint distributions that can be realized by the strategies $S$ is
$$ q(S) = \{ q \in \Delta({\aee}) : q(s) = q \ \text{for some} \ s \in S \}. $$
 
\subsection{Informal algorithm description}\label{s:algorithm}

The forthcoming algorithm is reminiscent of the trial and error learning algorithm introduced in \cite{Young2009} and can be viewed at a high level through the following diagram.
\begin{figure}[H]
\begin{center}
\includegraphics[trim = 0mm 100mm 20mm 40mm, clip, width=\columnwidth]{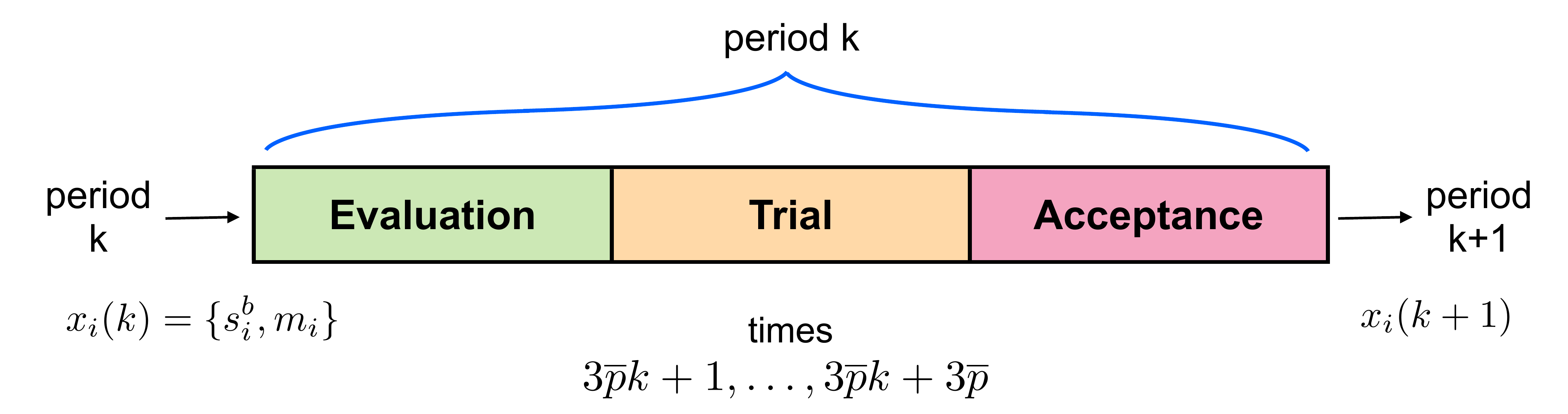}
\caption{Learning algorithm phases within each time period}
\end{center}
\end{figure}
The times $\{1, 2, \dots\}$ will be broken up into periods of length $3 \bar{p}$ where $\bar{p} > 1$ is an interval whose length will be defined formally below.  At the beginning of each period $k$, each agent $i \in N$ has a local state variable of the form $x_i(k) = [s_i^b, m_i]$ where $s_i^b \in S_i$ is the agent's baseline strategy and $m_i$ is the agent's mood.  The agent's baseline strategy corresponds to the strategy the agent is accustomed to playing.  The agent's mood $m_i$, which can either be \textsc{Content} or \textsc{Discontent}, dictates how likely each agent is to select its baseline strategy during a given period.  Roughly speaking, a content agent is more likely to select its baseline strategy while a discontent agent is more likely to try an alternate strategy.  

Each period $k > 0$, which consists of the time steps $\{3 \bar{p} k + 1, \dots, 3 \bar{p} (k + 1)\}$, will be broken up into three distinct phases called \emph{evaluation}, \emph{trial}, and \emph{acceptance}.  The behavior of the agents in each of these phases is highlighted below:

\vspace{.1cm}
\noindent -- \emph{Evaluation Phase:} The first phase is the \emph{evaluation phase}. In this phase, each agent  establishes a baseline utility, $u_i^b$, associated with its current baseline strategy, $s_i^b.$ All agents commit to their baseline strategies during this entire phase.

\vspace{.1cm}
\noindent -- \emph{Trial Phase:} The second phase is the \emph{trial phase}. During this phase, each agent has the opportunity to experiment with an alternate trial strategy, $s_i^t,$ in order to determine whether changing its baseline strategy could be advantageous. An agent's mood determines how likely it is to experiment.  In particular, a content agent will use its baseline strategy $s_i^b$ during the trial phase with high probability.  On the other hand, a discontent player is likely to experiment with a trial strategy $s_i^t \neq s_i^b$.  The exact probabilities associated with this selection process will be described in detail in the forthcoming section.   

\vspace{.1cm}
\noindent -- \emph{Acceptance Phase:} The third phase is the \emph{acceptance phase}. Here, an agent who experimented during the trial phase decides whether to accept its trial strategy or revert to its baseline strategy. Agents who did not experiment during the trial phase commit to their baseline strategies and observe payoff changes which occur due to others' changes in strategy.

\subsection{Formal algorithm description}\label{s:algorithm}

We begin by defining a constant $c > n$, an experimentation rate $\eps\in (0,1)$, and the length of a phase to be $\p = \lceil 1/\delta^{nc+1}\rceil$ time steps, for some small $\delta\in (0,1)$. A period consists of the evaluation, trial, and acceptance phases, and hence is $3\bar{p}$ time steps long. Let $x_i = x_i(k) = [s_i^b, m_i]$ represent that state of each agent $i \in N$ at the beginning of some period $k \in \{1, 2, \dots\}$.  We will formally present the algorithm using the same general structure given in previous section.  

\vspace{.2cm}

\noindent {\textbf{Agent Dynamics:}}  Here we describe how individual agents make decisions within a given period.  Decisions of an agent $i \in N$ are influenced purely by its state at the beginning of the $k$-th period, $x_i(k)$, and by payoffs received during the $k$-th period.  We specify agents' behavior during the $k$-th period for the three phases highlighted above.

\vspace{.2cm}
\noindent \emph{\textbf{-- Evaluation Phase:}} The evaluation phase consists of the times $t \in \{3\p k + 1, \dots, 3\p k+\p\}$.  Throughout this phase, each agent commits to its baseline strategy $s_i^b$.  At the end of the phase, each agent computes its average baseline utility, 
\begin{equation}\label{e:baseline payoff}
u_i^b = {1\over \p}\sum_{\tau = 3 \p k+1}^{3\p k+\p}U_i\bigl(\s_1^b(z(\tau), \dots, \s_n^b(z(\tau))  \bigr),
\end{equation}
where $z(\tau)$ denotes the common random signal observed at time $\tau$.  Here, $u_i^b$ is viewed as an assessment of the performance associated with the baseline strategy $s_i^b$.

\vspace{.1cm}
\noindent \emph{\textbf{-- Trial Phase:}} After the evaluation phase comes the trial phase which consists of the times $t \in \{(3\p k + \p)+1,\ldots ,3\p k +2 \p\}$.  During the trial phase each player $i\in N$ may try a strategy other than its baseline, and must commit to this trial strategy, $s_i^t \in S_i$, over the entire phase.  Agents' trial strategies are selected according to the following rule:

\begin{itemize}[leftmargin = .3cm]
%
\item \emph{Content, $m_i = C$:}
When agent $i$ is content, its trial strategy, $s_i^t \in S_i$, is chosen according to the  distribution
\begin{equation}
\Pr\left[s_i^t = s_i \right] = \left\{
\begin{array}{ll}
1 - \eps^c &\text{if }s_i = s_i^b \\
\eps^c \mathbin{/} |\mathcal{A}_i|&\text{for any } s_i = a_i\in\mathcal{A}_i
\end{array}\right.
\end{equation}
A strategy $s_i^t = a_i$ means that agent $i$ commits to playing action $a_i$ for the entire trial phase of the $k$-th period, i.e., the strategy does not depend on the common random signal.  Observe that a content player predominantly selects its baseline strategy during the trial phase.

\item \emph{Discontent, $m_i = D$:}
When agent $i$ is discontent, its trial strategy, $s_i^t$, is chosen randomly from the set $S_i$,
\begin{equation}\label{eq:981}
\Pr\left[s_i^t = s_i\right] = 1\mathbin{/} |S_i| \text{ for all } s_i\in S_i.
\end{equation}
%

%
\end{itemize}

At the end of the trial phase, each agent computes its average utility:
\begin{equation}\label{e:trial payoff}
u_i^t = {1\over \p}\sum_{\tau = 3\p k + \p)+1}^{3\p k +2 \p}U_i\bigl(\s_1^t(z(\tau), \dots, \s_n^t(z(\tau))  \bigr).
\end{equation}
Here, $u_i^t$ is viewed as an assessment of the performance associated with the baseline strategy $s_i^t$.  
\noindent \emph{\textbf{-- Acceptance Phase:}} The last phase is the acceptance phase which consists of times $t \in \{(3 \p k + 2\p)+1, \dots, 3\p k + 3\p\}$.  The primary purpose of the acceptance phase is to further evaluate changes in the payoffs between $u_i^b$ and $u_i^t$.  Each agent $i \in N$ commits to an acceptance strategy, denoted by $s_i^a \in S_i$, over the entire acceptance phase.  Each agent's acceptance strategy is selected according to the following.

\begin{itemize}[leftmargin = .3cm]
\item \emph{Content, $m_i = C$:}
When agent $i$ is content, its acceptance strategy is chosen as follows:
\begin{equation}
s_i^a = \left\{
\begin{array}{ll}
s_i^t &\text{if } u_i^t > u_i^b + \delta, \\
s_i^b &\text{if } u_i^t \leq u_i^b + \delta. 
\end{array}\right.
\end{equation}
That is, players only repeat their trial strategy if their performance was high enough relative to the performance of the baseline strategy.  

\item \emph{Discontent, $m_i = D$:}
When agent $i$ is discontent, the acceptance strategy is set as $s_i^a = s_i^t$.
\end{itemize}

Following the acceptance phase, each agent computes its average utility:
\begin{equation}\label{e:acceptance payoff}
u_i^a = {1\over \p}\sum_{\tau = (3 \p k + 2\p)+1}^{3\p k + 3\p}U_i\bigl(\s_1^a(z(\tau), \dots, \s_n^a(z(\tau))  \bigr).
\end{equation}
Here, $u_i^a$ is viewed as an assessment of the performance associated with the baseline strategy $s_i^a$.  

\vspace{.2cm}

\noindent {\textbf{State Dynamics:}}  After the agent dynamics comes the state dynamics which specifies how the state of each agent evolves.  The state of each agent $i \in N$ at the beginning of the $k+1$-st stage, i.e., $x_i(k+1)$, is influenced purely its state at the beginning of the $k$-th period, i.e., $x_i(k)$, the strategies $s_i^b,$ $s_i^t$ and $s_i^a$, and the payoffs received during the $k$-th period.   The state dynamics are broken into the following cases:

\vspace{.2cm}

\noindent \emph{{-- Content and No Experimentation, $m_i = C, s_i^t = s_i^b$:}} 
If agent $i$ was content at the start of the $k$-th period and did not experiment in the trial phase, its state at the beginning of the $(k+1)$-st period is chosen as follows:
\begin{itemize} [leftmargin = .3cm]
\item If $u_i^a \geq u_i^b - \delta,$
\begin{equation}\label{e:state trans1a}
x_i(k+1) = \left\{
\begin{array}{ll}
\left[s_i^a = s_i^b,\, C\right] &\quad\text{w.p. }1-\eps^{2c}, \\
\left[s_i^a = s_i^b,\, D\right] &\quad\text{w.p. } \eps^{2c}.
\end{array}\right.
\end{equation}
\item If $u_i^a < u_i^b - \delta$,
\begin{equation}\label{e:state trans1b}
x_i(k+1) = 
\left[s_i^a = s_i^b,\, D\right] 
\end{equation}
\end{itemize}
Accordingly, if the agent's average payoff during the acceptance phase is low enough, then it will become discontent.  

\vspace{.1cm}

\noindent \emph{{-- Content and Experimentation, $m_i = C, s_i^t \neq s_i^b$:}} 
If agent $i$ was content at the start of the $k$-th period and experimented during the trial phase, its state at the beginning of the $(k+1)$-st period is chosen as 
\begin{equation}\label{e:state trans2}
x_i(k+1) = \left[s_i^a, C\right].
\end{equation}
In this case the agent's average payoff during the acceptance phase does not impact its underlying state dynamics.  

\vspace{.1cm}

\noindent \emph{{-- Discontent, $m_i = D$:}} 
If agent $i$ was discontent at the start of the $k$-th period, its state at the beginning of the $(k+1)$-th period is chosen as follows
\begin{equation}\label{e:D state trans}
x_i(k+1) = \left\{
\begin{array}{ll}
\left[s_i^a, C\right] &\text{w.p. }  \eps^{1-u_i^a}, \\
\left[s_i^a, D\right] &\text{w.p. } 1 - \eps^{1-u_i^a}. 
\end{array}\right.
\end{equation}
Here, the agents are more likely to become content with strategies the yield higher average payoffs.

\subsection{Main Result}

Throughout this paper we focus on games where there is some degree of coupling between the utility functions of the agents.  The following definition of interdependence, taken from \cite{Young2009}, captures this notion of coupling.  
 
\begin{defn}
A game $G$ with agents $N = \{1,2,\ldots,n\}$ is said to be \emph{interdependent} if, for every $a\in\mathcal{A}$ and every proper subset of agents $J\subset N$, there exists an agent $i\notin J$ and a choice of actions $a_J^\prime\in\prod_{j\in J} \mathcal{A}_j$ such that $U_i(a_J^\prime,a_{-J})\neq U_i(a_J,a_{-J}).$
\end{defn}

Roughly speaking, the definition of interdependence states that it is not possibly to partition the group of agents into two sets whose actions do not impact one another's payoffs.

The following theorem characterizes the limiting behavior associated with the proposed algorithm. 

\begin{theorem}\label{t:main theorem}
Let $G = \left (N,\{U_i\},\{\mathcal{A}_i\}\right)$ be a finite interdependent game. 
First, suppose $q(S) \cap {\rm CCE} \neq \emptyset$.  Given any probability $p < 1$, if the exploration rate $\eps$ is sufficiently small, and if $\delta = \eps$, then for all sufficiently large times $t$,\footnote{For the proof of Theorem~\ref{t:main theorem}, we require $\delta = \eps$. However, in practice, fixing $\delta>\eps$ in order to shorten the period length, $\bar{p},$ often yields similar results, as we demonstrate in Example~\ref{e:example}.}
$$ {\rm Pr} \left[q(s(t)) \in \underset{ q \in q(S) \cap {\rm CCE}}{\arg \max} \ \sum_{i \in N} \sum_{a \in \aee} \ U_i(a) q^a \right] > p. $$ 
Alternatively, suppose $q(S) \cap {\rm CCE} = \emptyset$.  Given any probability $p < 1$, if the exploration rate $\eps$ is sufficiently small and $\delta = \eps$, then for all sufficiently large times $t$, 
$$ {\rm Pr} \left[q(s(t)) \in \underset{ q \in q(S)}{\arg \max} \ \sum_{i \in N} \sum_{a \in \aee} \ U_i(a) q^a \right] > p. $$ 
\end{theorem}

\vspace{.1cm}

We prove Theorem~\ref{t:main theorem} in Appendix~\ref{s:proof}.

A few remarks are on order regarding Theorem~\ref{t:main theorem}.  First, observe that the proposed algorithm is of the form (\ref{eq:2321}).  Second, the condition $q(S) \cap {\rm CCE} \neq \emptyset$ implies the agents can realize specific joint distributions that are coarse correlated equilibria through the joint strategy set $S$.  When this is the case, the above algorithm ensures the agents predominantly play a strategy $s \in S$ where the resulting joint distribution $q(s)$ corresponds to the efficient coarse correlated equilibrium.  Alternately, the condition $q(S) \cap {\rm CCE} = \emptyset$ implies there are no agent strategies that can characterize a coarse correlated equilibrium.  When that is the case, the above algorithm  ensures the agents predominantly play strategies that have full support on the action profiles $a \in \aee$ that maximize the sum of the agents payoffs, i.e., $\arg \max_{a \in \aee} \sum_{i \in N} U_i(a)$.

\subsection{Illustrative Example}\label{s:example}

Here, we present an example where agents update their strategies according to the algorithm above, and their actions converge to an efficient coarse correlated equilibrium.

\begin{example}\label{e:example}
Consider a game with two players, (Row, Column), and the following payoff matrix:
\begin{center}
\begin{tabular}{c|c|c|c|}
\multicolumn{1}{r}{}&
	\multicolumn{1}{c}{{L}}&
		\multicolumn{1}{c}{{M}}&
			\multicolumn{1}{c}{{R}}\\
\cline{2-4}T &0, 0&0, 1&0.85, 0.75\\
\cline{2-4}{M}&1, 0&0, 0&0, 0\\
\cline{2-4}{B}&0.75, 0.85&0, 0&0, 0\\\cline{2-4}
\end{tabular}
\end{center}

\smallskip

The efficient coarse correlated equilibrium in this game places probability 0.5 on joint action (T,R), and probability 0.5 on joint action (B,L), i.e., 
\begin{equation}\label{e:efficient CCE}
q^{(T,R)} = q^{(B,L)} = 0.5,
\end{equation}
and $q^a = 0$ for $a\notin\{(T,R), (B,L)\}.$ The expected utility associated with this coarse correlated equilibrium is 
$U_i(q) = 0.8.$

For each value of $\eps$ in $\{0.15, 0.1, 0.015,0.01\}$, we simulated our algorithm for 20 times over $10^5$ iterations, fixing $\delta = 0.14$. The table below shows the percentage of the last $5\times 10^4$ iterations spent in the efficient coarse correlated equilibrium as in \eqref{e:efficient CCE}.\footnote{We did not simulate our algorithm for smaller values of $\eps$ because convergence rates slow significantly as $\eps\to 0,$ reducing the algorithm's practicality. Next research steps include improving this algorithm's convergence rates. }

\begin{table}[H]
\begin{center}
\begin{tabular}{|c|c|}\hline
$\m{\eps}$ 	& \tb{\% time in efficient CCE}	\\\hline\hline
0.15 			& 9\%					\\\hline
0.1 			&16\%					\\\hline
0.015			&84\%					\\\hline
0.01			&87\%\\\hline
\end{tabular}
\end{center}
\end{table}%

Note that as $\eps$ decreases, more time is spent in the efficient coarse correlated equilibrium, as predicted by Theorem~\ref{t:main theorem}.



\end{example}

\section{Conclusion}

The majority of distributed learning literature has focused on identifying learning rules that converge to Nash equilibria.  However, alternate forms of behavior, such as correlated equilibrium, can often lead to significant improvements in system-wide behavior.  This paper focuses on identifying learning rules that converge to joint distributions that do not necessarily constitute Nash equilibria.  In particular, we have a provided a distributed learning rule, similar in spirit to the learning rule in \cite{Marden2013}, that ensuers agents play strategies that constitute efficient coarse correlated equilibria. A mild variant of the proposed algorithm could also ensure the agents play strategies that constitute correlated equilibria, as opposed to coarse correlated equilibria.  Future work seeks to investigate the applicability of such algorithms in the context of team versus team zero-sum games.

\bibliographystyle{plain}
\bibliography{CoarseCorrelatedEquilibria}

\begin{appendix}


\section{Proof of Theorem~\ref{t:main theorem}}

The formulation of the decision making process defined in Section~\ref{s:learning algorithm} ensures that the evolution of the agents' states over the periods $\{0, 1, 2, \dots\}$ can be represented as a finite ergodic Markov chain over the state space
\begin{equation}
X = X_1 \times \dots \times X_n
\end{equation}
where $X_i = S_i\times \{C,D\}$ denotes the set of possible states of agent $i$.  Let $P^\eps$ denote this Markov chain for some $\eps > 0$, and $\delta = \eps$.  Proving Theorem~\ref{t:main theorem} requires characterizing the stationary distribution of the family of Markov chains $\{P^\eps\}_{\eps > 0}$ for all sufficiently small $\eps$.  We employ the theory of resistance trees for regular perturbed processes, introduced in \cite{Young1993}, to accomplish this task.  We begin by reviewing this theory and then proceed with the proof of Theorem~\ref{t:main theorem}.

\subsection{Background: Resistance Trees}\label{a:resistance trees}

Define $P^0$ as the transition matrix for some nominal Markov process, and let $P^{\eps}$ be a perturbed version of this nominal process where the size of the perturbation is $\eps > 0$.  Throughout this paper, we focus on the following class of Markov chains.  

\begin{defn}\label{d:RPP}
A family of Markov chains defined over a finite state space $X$, whose transition matrices are denoted by $\{P^\eps\}_{\eps > 0}$, is called a \emph{regular perturbed process} of a nominal process $P^0$ if the following conditions are satisfied for all $x,x^\prime\in X$:
\begin{enumerate}
\item There exists a constant $c>0$ such that $P^\eps$ is aperiodic and irreducible for all $\eps \in (0,c]$.
\item $\lim_{\eps\to 0} P^{\eps}_{x \rightarrow x'}= P^0_{x \rightarrow x'}$.
\item If $P^\eps_{x \rightarrow x'} > 0$ for some $\eps>0$, then there exists a constant $r(x \to x') \geq 0$ such that 
\begin{equation}\label{e:RPP bounds}
0<\lim_{\eps\to 0}
\frac{P^\eps_{x \to x'}}
{\eps^{r(x \to x')}}<\infty.
\end{equation}
The constant $r(x \to x')$ is referred to as the \emph{resistance} of the transition $x \to x'.$
\end{enumerate}
\end{defn}

For any $\eps > 0$, let $\mu^{\eps} = \{\mu^{\eps}_x \}_{x \in X}  \in \Delta(X)$ denote the unique stationary distribution associated with $P^{\eps}$.  The theory of resistance trees presented in \cite{Young1993} provides efficient mechanisms for computing the support of the limiting stationary distribution, i.e., $\lim_{\eps \rightarrow 0^+} \mu^{\eps}$, commonly referred to as the stochastically stable states.  

\begin{defn}\label{d:ss}
A state $x \in X$ is \emph{stochastically stable} \cite{Foster1990} if $\lim_{\eps\to 0^+}\mu_x^\eps>0$, where $\mu^\eps$ is the stationary distribution corresponding to $P^\eps.$
\end{defn}

In this paper, we adopt the technique provided in \cite{Young1993} for identifying the stochastically stable states through a graph theoretic analysis over the recurrent classes of the unperturbed process $P^0$.  To that end, let $Y_0, Y_1, \dots, Y_m$ denote the recurrent classes of $P^0$.  Define ${\cal P}_{i j}$ to be the set of all paths connecting $Y_i$ to $Y_j$, i.e., a path $p \in {\cal P}_{i j}$ is of the form $p=\{(x_1, x_2), (x_2, x_3), \dots, (x_{k-1}, x_k)\}$ where $x_1 \in Y_i$ and $x_k \in Y_j$.  The resistance associated with transitioning from $Y_i$ to $Y_j$ is defined as 
\begin{equation}\label{eq:321}
r(Y_i , Y_j) = \min_{p \in {\cal P}_{i j}} \sum_{(x,x') \in p} r(x,x'). 
\end{equation}

The recurrent classes $Y_0,Y_1,\ldots,Y_m$ satisfy the following properties: (i) there is a zero resistance path, i.e., a sequence of transitions each with zero resistance, from any state $x \in X$ to at least one state $y$ in one of the recurrent classes; (ii) for any recurrent class $Y_i$ and any states $y_i,y_i' \in Y_i$, there is a zero resistance path from $y_i$ to $y_i'$; and (iii) for any state $y_i \in Y_i$ and $y_j \in Y_j$, $Y_i \neq Y_j$, any path from $y_i$ to $y_j$ has strictly positive resistance.  

The first step in identifying the stochastically stable states is to identify the resistance between the various recurrent classes.  %
The second step focuses on analyzing spanning trees of the weighted, directed graph $\mathcal{G}$ whose vertices are recurrent classes of the process $P^0,$ and whose edge weights are given by the resistances between classes in (\ref{eq:321}). Denote $\mathcal{T}_{i}$ to be the set of all spanning trees of $\mathcal{G}$ rooted at recurrent class $Y_i$. Next, we compute the stochastic potential of each recurrent class which is defined as follows:

\begin{defn}
The \emph{stochastic potential} of recurrent class $Y_i$ is 
\begin{equation*}
\gamma(Y_i) = \min_{{\cal T} \in \mathcal{T}_{i}} \sum_{(Y, Y')\in {\cal T}} r(Y,Y')
\end{equation*}
\end{defn}
The following theorem characterizes the recurrent classes that are stochastically stable.

\begin{theorem}[\cite{Young1993}] \label{t:Young Theorem}
Let $P^0$ be the transition matrix for a stationary Markov process over the finite state space $X$ with recurrent communication classes $Y_1,\ldots,Y_m$. For each $\eps > 0$, let $P^\eps$ be a regular perturbation of $P^0$ with a unique stationary distribution $\mu^\eps$. Then:
\begin{enumerate}
\item  As $\eps\to 0$, $\mu^\eps$ converges to a stationary distribution $\mu^0$ of $P^0.$
\item A state $x \in X$ is stochastically stable if and only if $x$ is contained in a recurrent class $Y_j$ that minimizes $\gamma(Y_j).$
\end{enumerate}
\end{theorem}

\subsection{Proof of Theorem~\ref{t:main theorem}}\label{s:proof}

We begin by restating the main results associated with Theorem~\ref{t:main theorem} (setting $\delta = \eps$) using the terminology defined in the previous section.  
\begin{itemize}[leftmargin = .3cm]
\item If $q(S) \cap {\rm CCE} \neq \emptyset$, then a state $x=\{x_i = [s_i, m_i] \}_{i \in N}$ is stochastically stable if and only if (i) $m_i = C$ for all $i \in N$ and (ii) the strategy profile $s = (s_1, \dots, s_n)$ constitutes an efficient coarse correlated equilibrium, i.e., 
\begin{equation}\label{e:SS1}
q(s) \in \underset{ q \in q(S) \cap {\rm CCE}}{\arg \max} \ \sum_{i \in N} \sum_{a \in \aee} \ U_i(a) q^a. 
\end{equation}
\item If $q(S) \cap {\rm CCE} = \emptyset$, then a state $x=\{x_i = [s_i, m_i] \}_{i \in N}$ is stochastically stable if and only if (i) $m_i = C$ for all $i \in N$ and (ii) the strategy profile $s = (s_1, \dots, s_n)$ constitutes an efficient action profile, i.e., 
\begin{equation}\label{e:SS2} 
q(s) \in \underset{ q \in q(S)}{\arg \max} \ \sum_{i \in N} \sum_{a \in \aee} \ U_i(a) q^a. 
\end{equation} 
\end{itemize}

\noindent For convenience, and with an abuse of notation, define
\begin{equation}
U_i(s) := \sum_{a\in \mathcal{A}}U_i(a)q^a(s)
\end{equation}
to be agent $i$'s expected utility with respect to distribution $q(s)$, where $s\in S.$

The proof of Theorem~\ref{t:main theorem} will consist of the following steps:
\begin{enumerate}[(i)]
\item Define the unperturbed process, $P^0$.
\item Determine the recurrent classes of process $P^0$.   
\item Establish transition probabilities of process $P^\eps$.
\item Determine the stochastically stable states of $P^\eps$ using Theorem~\ref{t:Young Theorem}.
\end{enumerate}

\vspace{.2cm}
\noindent \emph{Part 1:  Defining the unperturbed process}
\vspace{.2cm}

The unperturbed process $P^0$ is effectively the process identified in Section~\ref{s:learning algorithm} where $\eps = 0$.  Rather than dictate the entire process as done previously, here we highlight the main attributes of the unperturbed process that may not be obvious upon initial inspection.  

\begin{itemize}[leftmargin = .3cm]
\item If agent $i$ is content, i.e., $x_i = [s_i^b, C]$, the trial action is $s_i^t = s_i^b$ with probability $1$. Otherwise, if agent $i$ is discontent, the trial action is selected according to (\ref{eq:981}). 
\item The baseline utility $u_i^b$ in (\ref{e:baseline payoff}) associated with joint baseline strategy $s^b$ is now of the form
\begin{eqnarray}\label{eq:980}
u_i^b = U_i(s^b) .
\end{eqnarray}
This results from invoking the law of large numbers since $\p = \lceil 1/\eps^{nc+1}\rceil$.  The trial utility $u_i^t$ and acceptance utility $u_i^a$ are also of the same form.  
\item A content player will only become discontent if $u_i^a < u_i^b$ where associated payoffs are computed according to (\ref{eq:980}).
\end{itemize}

\vspace{.2cm}
\noindent \emph{Part 2: Recurrent classes of the unperturbed process}
\vspace{.2cm}

The second part of the proof analyzes the recurrent classes of the unperturbed process $P^0$ defined above.  The following lemma identifies the recurrent classes of $P^0$.  

\begin{lemma}\label{l:recurrent}
A state $x = (x_1,x_2,\ldots,x_n)\in X$ belongs to a recurrent class of the unperturbed process $P^0$ if and only if the state $x$ fits into one of following two forms:
\begin{itemize}[leftmargin = .3cm]
\item \emph{Form \#1:} The state for each agent $i \in N$ is of the form $x_i = \left[s_i^b,C\right]$ where $s_i^b \in S_i$. Each state of this form comprises a distinct recurrent classes.  We represent the set of states of this form by $C^0$.
\item \emph{Form \#2:} The state for each agent $i \in N$ is of the form $x_i = \left[s_i^b,D\right]$  where $s_i^b \in S_i$. All states of this form comprise a single recurrent class, represented by $D^0$. 
\end{itemize}
\end{lemma}
\vspace{.2cm}
\begin{proof}
We begin by showing that any state $x \in C^0$ is a recurrent class of the unperturbed process.  According to $P^0$, if the system reaches state $x$, then it remains at $x$ with certainty for all future time. Hence, each $x\in C^0$ is a recurrent class of $P^0.$  Next, we show that $D^0$ constitutes a single recurrent class.  Consider any two states $x,y\in D^0$.  According to the unperturbed process, $P^0$, the probability of transitioning from $x$ to $y$ is strictly positive $\left(\geq \prod_{i\in N}1/|S_i|\right)$; hence, the resistance of the transition $x \rightarrow y$ is $0$.  Further note that the probability of transitioning to any state not in $D^0$ is zero. Hence, $D^0$ forms a single recurrent class of $P^0$. 

The last part of the proof involves proving that any state $ x = \{[s_i^b, m_i]\}_{i \in N} \notin C^0 \cup D^0$ is not recurrent in $P^0$.  Since $x\notin {C^0\cup D^0}$, it consists of both content and discontent players.  Denote the set of discontent players by $J= \{i\in N\st m_i = D\} \neq \emptyset$.  We will show that the discontent players $J$ will play a sequence of strategies with positive probability that drives at least one content player to become discontent.  Repeating this argument at most $n$ times shows that any state $x$ of the above form will eventually transition to the all discontent state, proving that $x$ is not recurrent.  

To that end, let $x(1) = x$ be the state at the beginning of the $1$-st period.  According to the unperturbed process $P^0$, each discontent player randomly selects a strategy $s_i \in S_i$ which becomes part of the player's state at the ensuing stage.  Suppose each discontent agent selects a trial strategy $s_i = (a_i^1, \dots, a_i^w) \in {\cal A}_i^w \subset S_i$ during the $1$-st period, i.e., the discontent players select strategies of the finest granularization. Note that each agent selects a strategy with probability $\geq {1\mathop{/}|S_i|}.$  Here, the trial payoff for each player $i \in N$ associated with the joint strategies $s = (\{s_i^b\}_{i \notin J}, \{s_i\}_{i \in J})$ is 
\begin{eqnarray}
u_i^t(s) &=&  \int_{0}^{1} U_i(s(z)) dz \\
&=&  \frac{1}{w} U_i({a}) + \int_{w}^{1} U_i(s'(z)) dz, 
\end{eqnarray}
for some ${a} \in {\cal A}$ as $s_i(z) = s_i(z')$ for any $z,z' \in [0,1/w]$ for any agent $i \in N$.  If  $u_i^t < u_i^b$ for any any agent $i \notin J$, agent $i$ becomes discontent in the next stage and we are done.  

For the remainder of the proof suppose $u_i^t(s) \geq u_i^b(s^b)$ for all agents $i \notin J$. This implies all agents $N \setminus J$ will be content at the beginning of the second stage.  By interdependence, there exists a collective action $\tilde{a}_J \in \prod_{j \in J} {\cal A}_j$ and an agent $i \notin J$ such that $U_i(a) \neq U_i(\tilde{a}_J, a_{N\setminus J})$.   Suppose each discontent agent selects a trial strategy $s_i' = (\tilde{a}_i^1, a_i^2, \dots, a_i^w) \in {\cal A}_i^w \subset S_i$ during the second period, i.e., only the first component of the strategy changed.  The trial payoff for each player $i \in N$ associated with the joint strategies $s' = (\{s_i^b\}_{i \notin J}, \{s_i'\}_{i \in J})$ is 
\begin{eqnarray*}
u_i^t(s') &=&  \int_{0}^{1} U_i(s'(z)) dz \\
&=&  \frac{1}{w} U_i(\tilde{a}_J, a_{N\setminus J}) + \int_{w}^{1} U_i(s'(z)) dz \\ 
&\neq&  u_i^t(s)
\end{eqnarray*}
If $u_i^t(s') < u_i^t(s)$, agent $i$ will become discontent at the ensuing stage and we are done.  Otherwise, 
agent $i$ will stay content at the ensuing stage.  However, if each discontent agent selects a trial strategy $s_i'' = (a_i^1, a_i^2, \dots, a_i^w) \in {\cal A}_i^w \subset S_i$ during the third period, we know $u_i^t(s'') < u_i^t(s')$, where $s'' = (\{s_i^b\}_{i \notin J}, \{s_i''\}_{i \in J})$.  Hence, agent $i$ will become discontent at the beginning of period $4$. This argument can be repeated at most $n$ times, completing the proof.  
\end{proof}

\vspace{.2cm}
\noindent \emph{Part 3:  Transition probabilities of process $P^\eps$}
\vspace{.2cm}

Here, we establish the transition probability $P^\eps_{x\to x^+}$ for a pair of arbitrary states, $x,x^+\in X.$ 
 Let $x_i = [s_i,m_i]$, $x_i ^+= [s_i^+,m_i^+]$ for $i\in N,$  $s = (s_1,s_2,\ldots,s_n),$  and $s^+ = (s_1^+,s_2^+,\ldots,s_n^+).$ Then,
\begin{align}
P^\eps_{x\to x^+} 
&=\sum_{\tilde{s}s^t\in S}\sum_{\tilde{s}^a\in S}\biggl(\Pr[x^+\given s^t = \tilde{s}^t,\, s^a = \tilde{s}^a]\nonumber\\
&\hspace{.4in}\times\Pr[s^a = \tilde{s}^a \given s^t = \tilde{s}^t]\Pr[s^t = \tilde{s}^t]\biggr).\label{e:prob1}
\end{align}
Note that the strategy selections and state transitions are conditioned on state $x$; for notational brevity we do not explicitly write this dependence. Here, $s^t$ and $s^a$ represent the joint trial and acceptance strategies during the period before the transition to $x^+.$. The double summation in \eqref{e:prob1} is over all possible trial actions, $\tilde{s}^t\in S$, and acceptance strategies, $\tilde{s}^a\in S$. However, recall from \eqref{e:state trans1a} - \eqref{e:D state trans} that, when transitioning from $x$ to $x^+$, not all strategies can serve as intermediate trial and acceptance strategies. In particular, transitioning to state $x^+$ requires that $s^a = s^+;$ hence if $\tilde{s}^a\neq s^+,$ then
$\Pr[x^+\given s^t = \tilde{s}^t,\, s^a = \tilde{s}^a]=0,$ 
so we can rewrite \eqref{e:prob1} as:
\begin{align}
P^\eps_{x\to x^+}
& = \sum_{\tilde{s}^t\in S}\biggl(\Pr[x^+\given s^t = \tilde{s}^t,\, s^a = s^+]\nonumber\\
&\hspace{.3in} \times\Pr[ s^a = s^+\given s^t = \tilde{s}^t]\Pr[ s^t = \tilde{s}^t]\biggr)\label{e:prob2}
\end{align}
There are three cases for the transition probabilities in \eqref{e:prob2}. Before proceeding, we make the following observations. The last term in \eqref{e:prob2}, $\Pr[ s^t = \tilde{s}^t]$, is defined in Section~\ref{s:learning algorithm}; we will not repeat the definition here. For the first two terms, agents' state transition and strategy selection probabilities are independent when conditioned state $x$ and on the joint trial and acceptance strategy selections. Hence, we can write the first term as:
\begin{align}\label{e:first term}
\Pr[x^+\given s^t = \tilde{s}^t, s^a = s^+]= \prod_{i\in N}\Pr[x_i^+\given s^t = \tilde{s}^t, s^a = s^+]
\end{align}
and the second term as:
\begin{align}\label{e:second term}
\Pr[ s^a = s^+\given s^t = \tilde{s}^t] = \prod_{i\in N}\Pr[ s_i^a = s_i^+\given s^t = \tilde{s}^t].
\end{align}
The following three cases specify individual agents' probability of choosing the acceptance strategy $s_i^a$ in \eqref{e:second term} and transitioning to state $x_i^+$ in \eqref{e:first term}. 
 
\noindent\emph{Case (i) agent $i$ is content in state $x$, i.e., $m_i = C$, and did not experiment, $s_i^t = s_i$:}\\
\noindent For \eqref{e:second term}, since $s_i^a\in \{s_i^t,s_i\}$ we know that
\begin{align*}
\Pr[ s_i^a = s_i^+\given s^t = \tilde{s}^t]
=\left\{
\begin{array}{ll}
1 &\text{if } s_i^+ = s_i\\
0 & \text{otherwise}
\end{array}
\right..
\end{align*}
In \eqref{e:first term}, for any trial strategy $s^t = \tilde{s}^t$, the probability of transitioning to a state $x_i^+$ depends on realized average payoffs $u_i^b$ and $u_i^a$. In particular, if $x_i^+  = [s_i^+,C]$, then we must have that $u_i^a\geq u_i^b - \eps$, so
\begin{align*}
&\Pr\biggl[x_i^+ = [s_i^+,C]\given s^a = s^+, s^t = \tilde{s}^t\biggr]\\
&\hspace{.05in} = \int_0^1 \Pr[u_i^b = \eta ] \int_{\eta-\eps}^1 \Pr[u_i^a = \nu \given s^t = \tilde{s}^t, s^a = s^+] d\nu d\eta.
\end{align*}
Then, the probability that $x_i^+ = [s_i^+,D]$ is
$$1 - \Pr\biggl[x_i^+ = [s_i^+,C]\given s^a = s^+, s^t = \tilde{s}^t\biggr].$$

\noindent\emph{Case (ii) agent $i$ is content and experimented, $s_i^t\neq s_i:$}\\
\noindent For \eqref{e:second term}, agent $i$'s acceptance strategy depends on its average baseline and trial payoffs, $u_i^b$ and $u_i^t$. Recall, if $u_i^t\geq u_i^b+\eps,$ then $s_i^a = s_i$, i.e., agent $i$'s acceptance strategy is simply its baseline strategy from state $x$. Otherwise $s_i^a = s_i^t.$
Utilities $u_i^b$ and $u_i^t$ depend on joint strategies $s$ and $s^t$ and on the common random signals sent during the corresponding phases. Therefore, 
%
\begin{align*}
&\Pr[ s_i^a = s_i^+\given s^t = \tilde{s}^t\neq s]\nonumber\\
&\quad=\int_0^1\int_0^1 \Pr[ s_i^a = s_i^+\given u_i^b = \eta, u_i^t = \nu, s_i^t = s_i]\nonumber\\
&\hspace{.8in} \times \Pr[u_i^b = \eta]\Pr[u_i^t = \nu\given s^t = \tilde{s}^t]d\eta d\nu
\end{align*}
In \eqref{e:first term}, since agent $i$ remains content and sticks with its acceptance strategy from the previous period,
{\allowdisplaybreaks[3]\begin{align*}
&\Pr[x_i^+\given s^a = s^+, s^t = \tilde{s}^t]= \left\{
\begin{array}{ll}
1 & \text{if }s_i^+ = s_i^a\\
0 & \text{otherwise}
\end{array}
\right..
\end{align*}}

\noindent\emph{Case (iii) agent $i$ is discontent:} \\
\noindent For \eqref{e:second term}, 
\begin{align*}
\Pr[ s_i^a = s_i^+\given s^t = \tilde{s}^t]
= \left\{
\begin{array}{ll}
1 &\text{if } s_i^+ = s_i^t\\
0 &\text{otherwise}
\end{array}
\right. .
\end{align*}
In \eqref{e:first term}, agent $i$'s probability of becoming content depends only on its received payoff during the acceptance phase; it becomes content with probability $\eps^{1 - u_i^a}$ and remains discontent with probability $1 - \eps^{1 - u_i^a}$. Hence, if $x_i^+ = [s_i^+,C]$,
\begin{align*}
&\Pr\biggl[x_i^+ = [s_i^+,C] \given s^a = s^+, s^t = \tilde{s}^t\biggr]\\
&\quad= \int_0^1 \eps^{1 - \eta}\Pr[u_i^a = \eta \given s^a = s^+, s^t = \tilde{s}^t]d\eta.
\end{align*}
Then, 
\begin{align*}
&\Pr\biggl[x_i^+ = [s_i^+,D] \given s^a = s^+, s^t = \tilde{s}^t\biggr] \\
&\quad= 1 - \Pr\biggl[x_i^+ = [s_i^+,C] \given s^a = s^+, s^t = \tilde{s}^t\biggr]
\end{align*}

Now that we have established transition probabilities for process $P^\eps$, we may state the following lemma.
\begin{lemma} \label{l:RPP}The process $P^\eps$ is a regular perturbation of $P^0.$ 
\end{lemma}

It is straightforward to see that $P^\eps$ satisfies the first two conditions of Definition~\ref{d:RPP} with respect to $P^0$. The fact that transition probabilities satisfy the third condition, Equation \eqref{e:RPP bounds}, follows from the fact that the dominant terms in $P^\eps_{x\to y}$ are polynomial in $\eps$. This is immediately clear in all but the incorporation of realized utilities into the transition probabilities, as in \eqref{e:prob2}. However, for any joint strategy, $s$, and associated average payoff $u_i$, since 
$$\E[u_i] = \E\left[{1\over\bar{p}}\sum_{\tau = \ell}^{\ell+\bar{p}-1} U_i(\s(z(\tau)))\right] = U_i(\s).  $$
for any time period of length $\bar{p}$ in which joint strategy $s$ is played throughout the entire period. Moreover,  
$\Var\bigl[U_i(\s(z(\tau)))\bigr] \leq 1.$ Therefore, we may use Chebyschev's inequality and the fact that  $\bar{p} = \lceil  1\mathbin{/} \eps^{nc+2}  \rceil$ to see that
\begin{equation}\label{e:old claim}
\Pr \Bigl[\bigl| u_i - U_i(\s)\bigr|\geq \eps\Bigr] \leq { \Var\bigl[U_i(\s(z(\tau)))\bigr]\over \bar{p} \eps^2}\leq \eps^{nc}.
\end{equation} 
Note that this applies for all average utilities, $u_i^b, u_i^t,$ and $u_i^a$ in the aforementioned state transition probabilities.

%

\vspace{.2cm}
\noindent \emph{Part 3:  Determining the stochastically stable states}
\vspace{.2cm}



We begin by defining
\begin{align*}
&C^\star := \{x = \{[s_i,m_i]\}_{i\in N}\\
&\hspace{0.75in}\st q(s)\in \text{CCE} \text{ and } m_i = C,\,\forall i\in N\}\subseteq C^0
\end{align*} 
Here, we show that, if $C^\star$ is nonempty, then a state $x$ is stochastically stable if and only if $q(s)$  satisfies \eqref{e:SS1}. The fact that $q(s)$ must satisfy \eqref{e:SS2} when $C^\star = \emptyset$ follows in a similar manner. To accomplish this task, we (1) establish resistances between recurrent classes, and (2) compute stochastic potentials of each recurrent class.
\subsection*{Resistances between recurrent classes}



We summarize resistances between recurrent classes in the following claim. 

\begin{claim}\label{c:resistances}
 Resistances between recurrent classes satisfy:

\noindent For $x \in C^0$ with corresponding joint strategy $s$, 
\begin{equation}\label{e:D to x} r(D^0\to x) = \sum_{i\in N}(1 - U_i(\s)).\end{equation}



\noindent For a transition of the form $x\to y$, where $x\in C^\star$ and $y \in (C^0\cup D^0)\setminus \{x\},$ 
\begin{equation}r(x\to y)\geq 2c.\end{equation}

\noindent For a transition of the form $x\to y$  where $x\in C^0\setminus C^\star$ and $y\in (C^0\cup D^0 )\setminus \{x\}$,
\begin{equation}r(x\to y)\geq c.\end{equation}

\noindent For every $x\in C^0\setminus C^\star$, there exists a path 
$x = x^0\to x^1\to\cdots\to x^m\in C^\star\cup D^0$ with resistance 
\begin{equation}r(x^j\to x^{j+1}) = c,\;\forall j\in \{0,1,\ldots,m-1\}.\end{equation}

\end{claim}

These resistances are computed in a similar manner to the proof establishing resistances in \cite{Marden2013}; however, care must be taken due to the fact that there is a small probability that average received utilities fall outside of the window $U_i(s)\pm\eps$ during a phase in which joint strategy $s$ is played. We illustrate this by proving \eqref{e:D to x} in detail; the proofs are omitted for other types of transitions for brevity.

\begin{proof}
Let $x\in D^0$, $x^+\in C^0$ with $x_i = [s_i,D]$ and $x_i^+ = [s_i^+,C]$  for each $i\in N.$ Again, for notational brevity, we drop the dependence on state $x$ in the following probabilities. Note that all agents must select $s^t = s_i^+$ in order to transition to state $x_i = [s_i^+,C];$ otherwise the transition probability is 0.
we have
{\allowdisplaybreaks[3]\begin{align*}
&P^\eps_{x\to x^+} \nonumber\\
&\quad \stackrel{(a)}{=}\Pr[x^+\given s^a = s^+, s^t = s^+]\nonumber\\
&\hspace{1in} \times\Pr[ s^a = s^+\given s^t = s^+]\Pr[ s^t = s^+]\\
&\quad \stackrel{(b)}{=}\Pr[x^+\given s^a = s^+, s^t = s^+]\Pr[ s^t = s^+]\\
&\quad \stackrel{(c)}{=}\Pr[x^+\given s^a = s^+, s^t = s^+]\prod_{i\in N}1\mathop{/}|S_i|\\
&\quad = \prod_{i\in N}{1\over|S_i|}\Pr[x_i^+\given s^a = s^+,s^t = s^+]
\end{align*}}
where:
(a) follows from the fact that $s_i^a = s_i^t$ since $m_i = D$ in state $x$ for all $i\in N$,
(b) $\Pr[ s^a = s^+\given s^t = s^+] = 1$ since all agents are discontent and hence commit to their trial strategies during the acceptance period, and
(c) $\Pr[ s^t = s^+] = \prod_{i\in N}1\mathop{/}|S_i|$ since each discontent agent selects its trial strategy uniformly at random from $S_i$.

We now show that
\begin{equation}
0 < \lim_{\eps\to 0^+} \frac{P^\eps_{x\to x^+}}{\eps^{\sum_{i\in N}1 - U_i(s^+)}}<\infty
\end{equation}
satisfying \eqref{e:RPP bounds}. For notational simplicity, we define 
\begin{align}
U_i^+	&:= U_i(s^+)+\eps,\nonumber\\
U_i^-	&:=U_i(s^+) - \eps.\label{e:u defns}
\end{align}
We first lower bound $P^\eps_{x\to x^+} :$
\small
{\allowdisplaybreaks[3]\begin{align}
&P^\eps_{x\to x^+}	\nonumber\\
			&\quad= \prod_{i\in N}{1\over|S_i|}\Pr[x_i^+ \given s^a = s^+, s^t =s^+]\nonumber\\
			&\quad= \prod_{i\in N}{1\over |S_i|}\int_0^1 \Pr[u_i^a = \eta \given s^a = s^+, s^t = s^+]\eps^{1-\eta}d\eta\nonumber\\
			&\quad\geq \prod_{i\in N}{1\over |S_i|}\int_{U_i^-}^{U_i^+}\Pr[u_i^a = \eta\given s^a = s^+, s^t = s^+ ]\eps^{1-\eta}d\eta\nonumber\\
			&\quad\stackrel{(a)}{\geq}\prod_{i\in N}{\eps^{1 - U_i^-}\over |S_i|}\int_{U_i^-}^{U_i^+} \Pr[u_i^a = \eta\given  s^a = s^+, s^t = s^+] d\eta\nonumber\\
			&\quad\stackrel{(b)}{\geq}\prod_{i\in N}{\eps^{1 - U_i^-}\over |S_i|}(1-\eps^{nc})\nonumber\\
			&\quad={\eps^{\sum_{i\in N} 1 - U_i^-} + O(\eps^{nc})\over\prod_{i\in N} |S_i|}\label{e:lb}
\end{align}}
\normalsize
where
(a) is from the fact that $\eps^{1-\eta}$ is continuous and increasing in $\eta$ for $\eps\in(0,1),$ and 
(b) follows from \eqref{e:old claim}.  
Continuing in a similar fashion, it is straightforward to show 
\begin{equation}\label{e:ub}
P^\eps_{x\to x^+}\leq\eps^{\sum_{i\in N}(1-U_i^+)} + O(\eps^{nc}).
\end{equation}

Given \eqref{e:lb} and \eqref{e:ub}, and the fact that $U_i^+$ and $U_i^-$ satisfy \eqref{e:u defns}, we have that $P_{x\to x^+}^\eps$ satisfies \eqref{e:RPP bounds} with resistance $\sum_{i\in N}\left(1 - U_i(s^+)\right)$ as desired.
\end{proof}

\subsection*{Stochastic potentials}

The following lemma specifies stochastic potentials of each recurrent class. Using resistances from Claim~\ref{c:resistances}, the stochastic potentials follow from the same arguments as in \cite{Marden2013}. The proof is repeated below for completeness.

\begin{lemma}\label{l:sps}
Let $x\in C^0\setminus C^\star$ with corresponding joint strategy $s$, and let $x^\star\in C^\star$ with corresponding joint strategy $s^\star.$ The stochastic potentials of each recurrent class are:
\begin{align*}
\gamma(D^0) &= c|C^0\setminus C^\star| + 2c|C^\star|,\\
\gamma(x) &= \left(|C^0\setminus C^\star| - 1\right)c + 2c|C^\star| + \sum_{i\in N}(1 - U_i(\s)),\\
\gamma(x^\star) &= |C^0\setminus C^\star|c + 2c\left(|C^\star|-1\right) + \sum_{i\in N}(1 - U_i(\s^\star)),
\end{align*}
\end{lemma}

\emph{Proof:}
In order to establish the stochastic potentials for each recurrent class, we will lower and upper bound them.

\noindent\emph{Lower bounding the stochastic potentials}: To lower bound the stochastic potentials of each recurrent class, we determine the lowest possible resistance that a tree rooted at each of these classes may have.

\smallskip
\noindent 1) Lower bounding $\gamma(D^0)$:
$$\gamma(D^0) \geq c|C^0\setminus C^\star| + 2c|C^\star|$$ 
In a tree rooted at $D^0$, each state in $C^0$ must have an exiting edge. In order to exit a state in $C^0\setminus C^\star$, only a single agent must experiment, contributing resistance $c$. To exit a state in $C^\star$, at least two agents must experiment, contributing resistance $2c.$

\smallskip

\noindent 2) Lower bounding $\gamma(x)$, $x\in C^0\setminus C^\star$:
$$\gamma(x) \geq \left(|C^0\setminus C^\star| - 1\right)c + 2c|C^\star| + \sum_{i\in N}(1 - U_i(\s))$$ 
 Here, each state in $C^0\setminus \{x\}$ must have an exiting edge, which contributes resistance $\left(|C^0\setminus C^\star| - 1\right)c + 2c|C^\star|.$ The recurrent class $D^0$ must also have an exiting edge, contributing at least resistance $\sum_{i\in N}(1 - U_i(\s)).$

\smallskip

\noindent 3) Lower bounding $\gamma(x^\star)$, $x^\star\in C^\star$:
$$\gamma(x^\star) \geq |C^0\setminus C^\star|c + 2c\left(|C^\star|-1\right) + \sum_{i\in N}(1 - U_i(\s^\star))$$ 
 Again, each state in $C^0\setminus \{x^\star\}$ must have an exiting edge, which contributes resistance $\left(|C^0\setminus C^\star| - 1\right)c + 2c|C^\star|.$ The recurrent class $D^0$ must also have an exiting edge, contributing resistance at least $\sum_{i\in N}(1 - U_i(\s^\star)).$

\smallskip

\noindent\emph{Upper bounding the stochastic potentials:} In order to upper bound the stochastic potentials, we construct trees rooted at each recurrent class which have precisely the resistances established above.

\smallskip
\noindent 1) Upper bounding $\gamma(D^0)$:
$$\gamma(D^0) \leq c|C^0\setminus C^\star| + 2c|C^\star|$$ 
Begin with an empty graph with vertices $X$. For each state $x\in C^0\setminus C^\star$, add a path ending in $C^\star\cup D^0$ so that each edge has resistance $c$. This is possible due to Claim~\ref{c:resistances}. Now eliminate redundant edges; this contributes resistance at most $c|C^0\setminus C^\star|$ since each state in $C^0\setminus C^\star$ has exactly one outgoing edge. Finally, add an edge $x^\star \to D^0$ for each $x^\star\in C^0;$ this contributes resistance $2c|C^\star|$.

\smallskip

\noindent 2) Upper bounding $\gamma(x)$, $x\in C^0\setminus C^\star$:
$$\gamma(x) \leq \left(|C^0\setminus C^\star| - 1\right)c + 2c|C^\star| + \sum_{i\in N}(1 - U_i(\s)),$$ 
 This follows by a similar argument to the previous upper bound, except here we add an edge $D^0 \to x$ which contributes resistance $\sum_{i\in N}(1 - U_i(\s))$.

\smallskip

\noindent 3) Upper bounding $\gamma(x^\star)$, $x^\star\in C^\star:$
$$\gamma(x^\star) \leq |C^0\setminus C^\star|c + 2c\left(|C^\star|-1\right) + \sum_{i\in N}(1 - U_i(\s^\star)),$$ 
 This follows from an identical argument to the previous bound.
\hfill\QED

We now use Lemma~\ref{l:sps} to complete the proof of Theorem~\ref{t:main theorem}. For the first part, suppose $C^\star$ is nonempty, and let $$x^\star\in \argmax_{x\in C^\star} \sum{U_i(\s)},$$ where joint strategy $s$ corresponds to state $x$.  Then,
\begin{align*}
\gamma(x^\star)  	&=  |C^0\setminus C^\star|c + 2c\left(|C^\star|-1\right) + \sum_{i\in N}(1 - U_i(\s^*))\\
			&<  |C^0\setminus C^\star|c + 2c|C^\star|  \quad\text{(since $c\geq n$)} \\
			&= \gamma(D).
\end{align*}
For $x \in C^0$, 
\begin{align*}
\gamma(x^\star)  	&=  |C^0\setminus C^\star|c + 2c\left(|C^\star|-1\right) + \sum_{i\in N}(1 - U_i(\s^\star))\\
			&<   |C^0\setminus C^\star - 1|c + 2c\left(|C^\star|\right) + \sum_{i\in N}(1 - U_i(s)) \\
			&=\gamma(x).
\end{align*}
For $x\in C^\star$ with $$x\notin \argmax_{x\in C^\star} \sum{U_i(\s)},$$
\begin{align*}
\gamma(x^\star)  	&=  |C^0\setminus C^\star|c + 2c\left(|C^\star|-1\right) + \sum_{i\in N}(1 - U_i(\s^\star))\\
				&<  |C^0\setminus C^\star|c + 2c\left(|C^\star|-1\right) + \sum_{i\in N}(1 - U_i(s)\\
				&=\gamma(x).
\end{align*}
Applying Theorem~\ref{t:Young Theorem}, $x^\star$ is stochastically stable. Since all other states have strictly larger stochastic potential, \emph{only} states $x^\star\in C^\star$ with $x^\star\in \argmax_{x\in C^\star} \sum{U_i(\s)}$ are stochastically stable. From state $x^\star$, if each agent plays according to its baseline strategy, then the probability that joint action $a\in \mathcal{A}$ is played at any given time is $\Pr(a  = a^\prime) = q^{a^\prime(s^\star)}.$  This implies that a CCE which maximizes the sum of agents' payoffs is played with high probability as $\eps\to 0,$ after sufficient time has passed.

The second part of the theorem follows similarly by considering the case when $C^\star = \emptyset.$ \hfill\QED

\end{appendix}

\end{document}